# A Memadmittance Systems Model for Thin Film Memory Materials

Blaise Mouttet

George Mason University, VA, USA, bmouttet@gmu.edu

## ABSTRACT

In 1971 the memristor was originally postulated as a new non-linear circuit element relating the time integrals of current and voltage. More recently researchers at HPLabs have linked the theoretical memristor concept to resistance switching behavior of $TiO_{2-x}$ thin films. However, a variety of other thin film materials exhibiting memory resistance effects have also been found to exhibit a memory capacitance effect. This paper proposes a memadmittance (memory admittance) systems model which attempts to consolidate the memory capacitance effects with the memristor model. The model produces equations relating the cross-sectional area of conductive bridges in resistive switching films to shifts in capacitance.

## 1  INTRODUCTION

In 1971, Prof. Leon Chua of UC Berkeley published a paper [1] arguing that the conventional view of passive circuitry consisting of resistors, capacitors, and inductors was incomplete. A new fundamental circuit element called the "memristor" (memory resistor) was proposed as a fourth fundamental circuit element based on a functional relationship between the time integral of voltage ($\varphi=\int v(t)dt$) and the time integral of current ($q=\int i(t)dt$). Mathematically this may be expressed as:

$$\varphi = M(q) \qquad (1)$$

where M(*) is a continuous, non-linear function.

A subsequent paper [2] in 1976 by Chua and Sung-Mo Kang extended and formalized the theory to cover a broader range of systems characterized by a pinched, zero-crossing hysteresis curve which may mathematically be described by the coupled non-linear systems equations:

$$v(t) = R(w,i,t)i(t) \qquad (2a)$$

$$dw/dt = f(w,i,t) \qquad (2b)$$

where v(t) is voltage as a function of time t, i(t) is current as a function of time t, w is a state variable, R(*) is a state-dependent resistance function, and f(*) is a function representative of the dynamic change of the resistance state. It is notable that one special cases of a memristance system exists when R(w,i,t) = R(w) and dw/dt = 0 which corresponds to the case of a linear, time-invariant resistor. Another special case exists when R(w,i,t) = R(w) and dw/dt = i(t) which corresponds to the originally defined memristor. Thus memristive systems unify resistors and memristors under one comprehensive system.

In 2008 researchers at HPLabs published a paper [3] connecting the theoretical work of Chua and Kang to real material systems under the interpretation that the state variable (w) corresponded to a measure of ionic distribution in $TiO_2$ thin films and the change of state (dw/dt) was a measure of ionic drift. This model led to an expression of a memristive system in terms of the measured low resistance ($R_{ON}$) and measured high resistance ($R_{OFF}$) such that

$$v(t) = [R_{ON}(w/D)+R_{OFF}(1-w/D)]\, i(t) \qquad (3a)$$

$$dw/dt = \mu_v R_{ON}/D\, i(t) \qquad (3b)$$

where (D) was used to represent the film thickness and $\mu_v$ was used to represent the average ion mobility of the film. It is notable that (although not cited in the HPLab paper) the state-dependent resistance switching property of thin film $TiO_2$ was experimentally realized 40 years earlier including data curves later associated with memristive systems [4].

Further papers have attempted to extend the memristive model of $TiO_2$ by incorporating an electric-field dependent ionic mobility in order to explain the non-linearity of the ionic drift rate with respect to the applied voltage [5] as well as providing an alternative model using the tunneling gap as the state variable (w) [6]. One issue that has not been carefully analyzed in these models is the capacitance of the system. Capacitance hysteresis effects are well recognized as a consequence of ionic trapping and ionic drift in the oxides of MOS systems and have been observed in a variety of thin film materials [7-11]. While the theoretical proposal of a "memcapacitor" has been put forward in analogy to the memristor [12] it has yet to be connected to material parameters of thin films in any specific detail. Section 2 develops an extended version of memristive systems equations applicable to thin films and including capacitance effects. Section 3 describes a potential application of the model to the analytical determination of

March 14, 2010

the length change and area of conductive bridges created during resistance switching.

## 2  MEMADMITTANCE SYSTEMS MODEL

Memristance effects have been attributed to an oxide layer having an ionic content and a nanoscale thickness by [3]. For a sufficiently thin ionic film a functional relationship would then exist between charge (q) and flux-linkage ($\varphi$). However, in the same material system having a larger oxide thickness it would be expected that a capacitive functional relationship exists between charge (q) and voltage (v). Thus it would be reasonable to expect both memristive and capacitive effects to co-exist in metal-insulator-metal thin films having thicknesses between the extreme of purely memristive at the nanometer thickness scale and purely capacitive at the micrometer thickness scale. To incorporate both effects (at least at a theoretical level) a functional relationship should exist such that the charge (q) of the system can be defined in terms of both the flux linkage ($\varphi$) (as in a memristor) and the voltage (v) (as in a capacitor).

$$q = q(\varphi,v) \qquad (4)$$

The current (i) of the system may be found by differentiating (5) with respect to time producing an equation analogous to that found for a parallel resistor and capacitor combination,

$$i(t) = G(\varphi,v)v(t)+C(\varphi,v)dv(t)/dt \qquad (5)$$

where $G(\varphi,v)$ and $C(\varphi,v)$ are conductance and capacitance functions defined by

$$G(\varphi,v) = \partial q(\varphi,v)/\partial \varphi \qquad (6)$$

$$C(\varphi,v) = \partial q(\varphi,v)/\partial v. \qquad (7)$$

It is notable that if the charge is a continuous function of the flux-linkage and voltage, the functional relationship of (6) and (7) imply a mathematical coupling between the conductance and capacitance functions such that:

$$\partial C(\varphi,v)/\partial \varphi = \partial G(\varphi,v)/\partial v \qquad (8)$$

In cases where the conductance and capacitance functions are both determined by a common state variable related to the ionic distribution in a thin film (5) may be rewritten in terms of non-linear state equations

$$i(t) = G(w)v(t)+C(w)dv(t)/dt \qquad (9)$$

$$dw/dt = f(w,v) \qquad (10)$$

in which f(*) is a function defining the voltage-dependent ionic drift.

March 14, 2010

Based on the analysis of a dual layer $TiO_{2-x}/TiO_2$ system discussed in [3] an insulator layer of thickness (D) is separated into a high conductive region of thickness (w) representing the $TiO_{2-x}$ material and a low conductive region of thickness (D-w) representing the dielectric $TiO_2$ material. In accordance with this analysis the conductance function G(w) may be expressed as the reciprocal of (3a).

$$G(w) = [R_{ON}(w/D)+R_{OFF}(1-w/D)]^{-1} \qquad (11)$$

For a uniform area distribution of $TiO_{2-x}$ in which the dielectric region thickness (D-w) is large enough to avoid breakdown a parallel plate capacitance model may be used to approximate a state-dependent capacitance as:

$$C(w) = \varepsilon A/(D-w) \qquad (12)$$

where ($\varepsilon$) is the permittivity of the dielectric $TiO_2$, (A) is the cross-sectional area of the metal-insulator-metal junction, and (D-w) is the thickness of the $TiO_2$ dielectric region.

A simple first-level model for the ionic drift (dw/dt) may be expressed as a linear function of the electric field within the junction. Two components contributing to the electric field in the dielectric region are the external electric field ($E_{ext}$) which is a function of the applied voltage ($V_a$) and the internal electric field ($E_{int}$) determined by the ionic distribution within the junction

$$dw/dt = \mu_v(E_{ext}(V_a)-E_{int}(w)) \qquad (13)$$

where $\mu_v$ is the ionic mobility. The application of the external field will push the ions toward one of the electrodes resulting in a build-up of ions which will generate an internal field in the opposite direction. It is notable that in various materials exhibiting memristive effects this can result in either an electrochemical reaction with the electrode material leading to the growth of a metallic filament from the electrode or an accumulation of defect sites forming a conductive bridge [13]. If the surface geometry of the electrodes exhibit any non-uniformity it is likely that the ion accumulation will be attracted to this non-uniformity and serve as a basis for the initial growth of the conductive bridge. Fig. 1 provides an illustration of such ionic accumulation leading to filament formation.

The external field ($E_{ext}$) in the dielectric region may be calculated from

$$E_{ext} = -(V_a - V_T)/(D-w) \qquad (14)$$

where $V_a$ is the applied voltage, $V_T$ denotes any threshold voltage which may exist due to the relative work functions of the metal used for the electrodes and the oxide, and (D-w) is the film thickness of the dielectric region. While (14) uses a singular value for the state variable (w) it is notable that in the more general case of an area-dependent boundary between the conductive and dielectric regions such as in Fig.1 the state variable would be expressed as a state function which varies across the cross-sectional area. The internal field ($E_{int}$) may be calculated based on Gauss's Law relating the electric displacement ($\varepsilon E_{int}$) and an ionic density function of the junction ($N_v$)

$$\nabla(\varepsilon E_{int}) = -qN_v(w) \quad (15)$$

where $\varepsilon$ is the permittivity of the dielectric film and $q$ is the electronic charge. It is noted that this is only a first-level linear model of the ionic drift equation. More advanced models should incorporate thermal diffusion effects in addition to non-linear effects occurring at high fields as explained in [5].

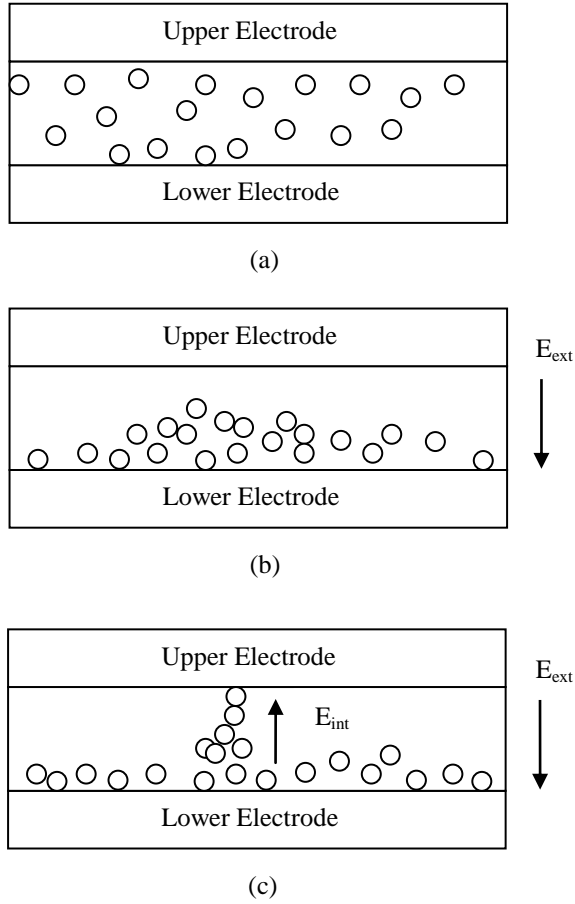

Figure 1: (a) Uniform distribution of oxygen vacancies or ions in thin film (b) External electric field ($E_{ext}$) initiates drift toward one electrode (c) A localized internal electric field is generated by attraction of ions to a non-uniform region of the electrode. The build-up of ionic density extends the internal field and promotes the creation of a conducting bridge.

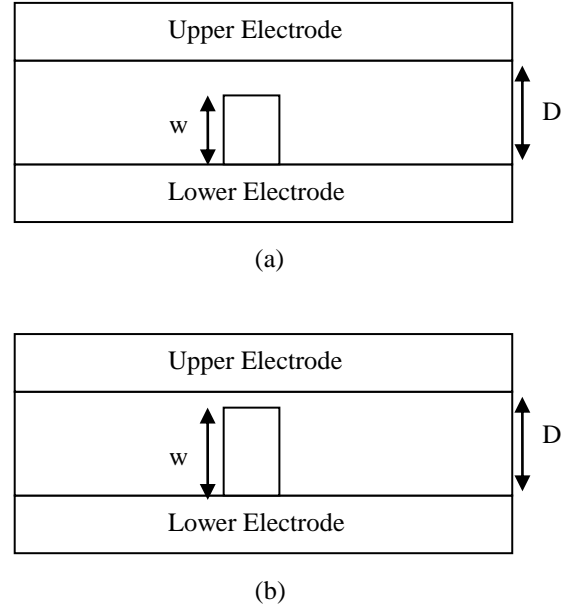

Figure 2: (a) High resistance state of junction in which (D-w) is sufficiently large so that there is minimum tunneling and the junction is primarily capacitive (b) Low resistance state of conductive bridge in which (D-w) is small enough to provide significant electron tunneling and the junction is primarily resistive.

## 3 RELATION TO MATERIAL PARAMETERS

For materials which exhibit both capacitance and resistance switching based on the formation of a conductive bridge the dimensions of the bridge may be estimated using the measurements of shifts in the resistance and capacitance .Fig. 2 illustrates a two state system having a junction switched between a capacitive state, in which the thickness of the dielectric gap is relatively large, and a resistance state, in which the gap is relatively small. In the OFF state of Fig.2a the junction resistance $R_{OFF}$ scales with the inverse of the tunneling current as

$$R_{OFF} = R_0/[\exp(-2\pi\sqrt{(8m_e\phi/h)}(D-w_{OFF}))] \quad (16)$$

where ($R_0$) is a proportionality constant, ($m_e$) is the electron effective mass, ($\phi$) is the tunneling potential barrier, ($h$) is Planck's constant, ($w_{OFF}$) is the length of the conductive bridge in the OFF state, and ($D-w_{OFF}$) is the tunneling gap in the OFF state. The total capacitance in the OFF state can be calculated based on the sum of the capacitance formed between the upper and lower electrode and the capacitance formed between the filament and the upper electrode.

$$C_{OFF} = \varepsilon(A-A_{cb})/D + \varepsilon A_{cb}/(D-w_{OFF}) \quad (17)$$

where ($\varepsilon$) is the permittivity constant of the dielectric, (A) is the electrode area, and ($A_{cb}$) is the conductive bridge tip area. In the ON state of Fig. 2(b) the tunneling resistance is proportional to the inverse of the tunneling current as:

$$R_{ON} = R_0/[\exp(-2\pi\sqrt{(8m_e\phi/h)}(D-w_{ON}))] \qquad (18)$$

where ($w_{ON}$) is the length of the conductive bridge in the ON state.

In the ON state with a relatively low ON resistance the current flow would eliminate charge storage in the bridge and thus the conductive bridge capacitance is negligible relative to the capacitance between the electrodes and the ON capacitance may be calculated as:

$$C_{ON} = \varepsilon(A - A_{cb})/D \qquad (19)$$

Using (16) and (18) the change in the conductive bridge length may be calculated as:

$$w_{ON} - w_{OFF} = [2\pi\sqrt{(8m_e\phi/h)}]^{-1} \ln(R_{OFF}/R_{ON}) \qquad (20)$$

Using (17) and (19) the change in capacitance is found to be related to the material parameters as:

$$C_{OFF} - C_{ON} = \varepsilon A_{cb}/(D - w_{OFF}) \qquad (21)$$

In materials having a negligible capacitance shift this may be indicative that the conductive bridge is a filamentary structure with a small tip area ($A_{cb}$). However, in systems which do exhibit a measureable capacitance shift this equation may be useful in approximating the conductive bridge tip area.

Some general predictions are made from this model as follows:

1) Memcapacitance effects will be found more pronounced in materials in which the conductive bridge area extends across the electrode area (i.e. $A_{cb}$ is large).

2) Asymmetries in the electrode geometry will promote localized filament formation and thus reduce memcapacitive effects. Conversely electrodes that are manufactured with a higher degree of planarity will be more likely to produce higher $A_{cb}$ values and a higher memcapacitive effect.

3) Materials having lower ionic mobilities will be more likely to produce memcapacitive effects since the lateral ionic drift will be smaller and thus the localization of the conductive bridge will be minimized. (This may be the reason why memory capacitance effects have been more noticeable in nanoparticle doped films [7,8,10] which are likely to reduce the mobility).

4) Materials exhibiting memcapacitive effects may be more slowly switching if due to lower ion mobilities.

March 14, 2010

5) Materials exhibiting memcapacitive effects may be more stable since a large conductive bridge area would likely require more energy to disrupt. Conversely it may require more energy to alter (i.e. erase) memcapacitive memory which is an issue for RRAM applications.